\documentclass[11pt]{article}
\usepackage{graphicx}
\usepackage{subfigure}
\usepackage{epsfig}
\usepackage{multirow}

\newcommand{\BABARPubYear}    {06}

\newcommand{\BABARConfNumber} {023}
\newcommand{\SLACPubNumber} {11989}
\newcommand{\LANLNumber} {0607045}

\input babarsym

\newcommand{\roots}        {\ensuremath{\sqrt{s}}\xspace}

\newcommand{\ntaupair}         {\ensuremath{279.8\times 10^6}\xspace}

\newcommand{\Mnu}             {\ensuremath{m_{\nu}^2}\xspace}

\newcommand{\tagmass}          {\ensuremath{M_{\mathrm{tag}}}\xspace}

\def\eff   {\ensuremath{\varepsilon}\xspace}

\def\DeltaE {\ensuremath{\Delta E}\xspace}

\def\tenseven {\ensuremath{\times 10^{-7}}\xspace}

\newcommand{\CM} {\mbox{C.M.}\xspace}
\newcommand{\gevccgevcc}{\ensuremath{{\mathrm{\,Ge\kern -0.1em V^2\!/}c^4}}\xspace}
\newcommand{\eett}   {\ensuremath{e^+e^- \to \tautau}\xspace}
\def\kk       {\mbox{\tt KK2F}\xspace}
\def\tauola     {\mbox{\tt TAUOLA}\xspace}

\def\koralb     {\mbox{\tt KORALB}\xspace}
\def\evtgen     {\mbox{\tt EVTGEN}\xspace}
\def\jetset     {\mbox{\tt JETSET}\xspace}
\def\geant     {\mbox{\tt GEANT4}\xspace}

\long\def\inst#1{\par\nobreak\kern 4pt\nobreak
    {\it #1}\par\vskip 10pt plus 3pt minus 3pt}

\begin{document}
{\pagestyle{empty}

\begin{flushright}
%BAD 1564 V12\\
\babar-CONF-\BABARPubYear/\BABARConfNumber \\
%\babar-PUB-\BABARPubYear/\BABARPubNumber \\
SLAC-PUB-\SLACPubNumber \\
hep-ex/\LANLNumber \\
\end{flushright}

\par\vskip 5cm

% Title of the paper
\begin{center}
\Large {\bf Study of the Lepton Flavor Violating Decay $\tau^{-} \rightarrow \mu^{-} \eta$}
\end{center}
\bigskip

\begin{center}
\large The \babar\ Collaboration\\
\mbox{ }\\
\today
\end{center}
\bigskip \bigskip

% Abstract
\begin{center}
\large \bf Abstract
\end{center}
We present the preliminary results from the search of the non-conservation of
lepton flavor number in the decay of a \mtau to a lighter mass lepton
and a pseudo-scalar meson, performed using $\epem\to\tautau$ events collected 
at a center-of-mass energy near 10.58\gev with the \babar\ detector at the SLAC PEP-II $\epem$ storage ring.
No evidence of such a signal has been found in the data sample corresponding to a luminosity of 314.5\invfb, 
and we set an upper limit of 1.6\tenseven at 90\% confidence level on the decay of $\tau^-\to\mu^-\eta$.
\vfill
\begin{center}

Submitted to the 33$^{\rm rd}$ International Conference on High-Energy Physics, ICHEP 06,\\
26 July---2 August 2006, Moscow, Russia.

\end{center}

\vspace{1.0cm}
\begin{center}
{\em Stanford Linear Accelerator Center, Stanford University, 
Stanford, CA 94309} \\ \vspace{0.1cm}\hrule\vspace{0.1cm}
Work supported in part by Department of Energy contract DE-AC03-76SF00515.
\end{center}

\newpage
} % end of pagestyle{empty}

% Input author list file
%
%author list removed temporarily to save trees 7/9/04 RNC
%
\begin{center}
\small

The \babar\ Collaboration,
\bigskip

%% author list as of 01-Jul-2006 (596 authors)
%
{B.~Aubert,}
{R.~Barate,}
{M.~Bona,}
{D.~Boutigny,}
{F.~Couderc,}
{Y.~Karyotakis,}
{J.~P.~Lees,}
{V.~Poireau,}
{V.~Tisserand,}
{A.~Zghiche}
\inst{Laboratoire de Physique des Particules, IN2P3/CNRS et Universit\'e de Savoie,
 F-74941 Annecy-Le-Vieux, France }
% ...

\end{center}\newpage

% The body of the paper starts here
\section{INTRODUCTION}
\label{sec:Introduction}

The search for lepton flavor violation (LFV) in charged lepton processes is one of the
cleanest ways to look for new physics beyond the Standard Model (SM). In the SM with the 
massless neutrinos, the total lepton number and lepton flavors are conserved, so that processes like
$\tau^{-} \to \mu^{-}\eta$~\cite{conjugate} are strictly forbidden. The recent discovery of neutrino oscillations~\cite{NuOsc}
indicates that the neutrinos have a very small and non-zero mass, and the lepton flavor is not conserved in the neutral lepton sector.
However, in the simplest extension of the SM, which accommodates 
the neutrino mass and mixing, the expected branching ratios for lepton flavor violating processes 
in the charged lepton sector are too small to be experimentally accessible.
Thus, any observation of LFV in the charged lepton sector would be an unambiguous signature of new physics
beyond the SM~\cite{Ma:2002pq}. The decay $\tau^{-} \to \mu^{-}\eta$ is of particular interest because it could be 
enhanced in supersymmetric models~\cite{sher} due to the potentially large coupling of the Higgs boson 
to the $s\bar{s}$ pairs and its associated color factors.
The most stringent upper limit published to date on this decay is $\BR_{UL} < 1.5\tenseven$ 
at 90\% confidence level (c.l.) with 154\invfb of $\epem$ annihilation data collected 
by the Belle experiment~\cite{enari2:2005xw}.

\section{THE \babar\ DETECTOR AND DATASET}
\label{sec:babar}

The results presented in this paper are based upon data collected by the \babar\ detector at the PEP-II storage ring. 
Details of the detector are described elsewhere~\cite{detector}. 
Charged particles are reconstructed as tracks with a 5-layer silicon vertex tracker (SVT) and a 40-layer drift chamber (DCH)
inside a 1.5-T solenoidal magnet. An electromagnetic calorimeter (EMC) consisting of 6580 CsI(Tl)
crystals is used to identify electrons and photons. 
A ring-imaging Cherenkov detector (DIRC) is used to identify
charged pions and kaons and provides additional electron identification information.
The flux return of the solenoid, instrumented with resistive plate chambers (IFR) and limited streamer tubes (LST), 
is used to identify muons.

The data sample consists of an integrated luminosity of \L = 314.5\invfb collected at a center-of-mass (\CM) energy 
\roots near 10.58\gev.
With an average cross section of $\sigma_{\eett}$ = (0.89$\pm$0.02) nb~\cite{kkxsec} as determined using the 
\kk Monte Carlo (MC) generator~\cite{kk2f},
this corresponds to a data sample of \ntaupair $\tau$-pair events.

The signal MC is generated with a two-body decay $\tau^- \to \mu^-\eta$ model,
incorporated into the \kk generator employing the \tauola decay package~\cite{tauola}.
The potential backgrounds including $\epem \to \mu^+\mu^-, \tautau, (\uubar, \ddbar, \ssbar$ mixture), and $\ccbar$ processes
were studied using the \kk, \evtgen~\cite{evtgen} \& \jetset~\cite{jetset} generators.
We also study $b\bar{b}$ Monte Carlo events but find that no events pass our selection criteria.
The detector response is modeled using the \geant simulation package~\cite{geant}.

\section{ANALYSIS METHOD}

\subsection{Reconstruction and event selection}

We reconstruct the signal process $e^+e^- \to \tau^+\tau^-$ with one $\tau^{-}$ decaying
to $\mu^{-}\eta$, where $\eta \to \gamma\gamma$ or $\pi^{+}\pi^{-}\piz$,
using events with zero total charge and with two or four well-reconstructed tracks,
where none of the tracks originate from conversion of photons in the material of the detector.
The event is divided into hemispheres by the plane perpendicular to the thrust axis~\cite{thrust},
which characterizes the direction of maximum energy flow in the event,
and is calculated using the energy-momentum information 
from all the tracks observed in the DCH and energy deposits in the EMC.

For the decay $\tau^{-} \to \mu^{-}\eta~(\eta \to \gamma\gamma)$, the signal-side hemisphere
is required to contain a pair of energy deposits in the EMC each consistent with being a photon above a 100\mev energy threshold~\cite{LAB}
and only one track (1-prong hemisphere). Extra energy deposits in the EMC greater than the 100\mev threshold 
unassociated with any of the identified \mtau daughters are not allowed in the signal-side hemisphere.

For the 3-prong decay $\tau^{-} \to \mu^{-}\eta ~(\eta \to \pi^{+}\pi^{-}\piz)$,
a pair of energy deposits in the EMC each consistent with being a photon above a 50\mev energy threshold 
is used to form the $\piz$ candidate.
We then form two possible $\eta$ candidates by combining the $\piz$ and the positively charged track with 
either of the two negatively charged tracks in the signal-side hemisphere.
The $\eta$ candidate closest to its nominal mass (547.75$\pm$0.12\mevcc~\cite{pdg04}) is assigned to be the true candidate,
and the negatively track not associated with the $\eta$, is assigned to be the signal muon candidate.

We require all possible $\eta$ candidates, constructed out of their respective daughters,
to have an energy greater than 1.4\gev and 1.2\gev for the decay $\eta \to \gamma\gamma$ and $\eta \to \pi^{+}\pi^{-}\piz$, respectively.
We then form the complete \mtau decay chain, and finally accept an event in either one of the decay modes of $\eta$ exclusively, 
based upon the closeness of the signal \mtau candidate to its nominal mass (1.777\gevcc~\cite{bes}).

For the 1-prong ($\eta \to \gamma\gamma$) signal candidates,
we assign the origin of the two photons from the $\eta$ to come from the point of closest approach 
of the signal lepton track to the $\epem$ collision axis.
For the 3-prong ($\eta \to \pi^{-}\pi^{-}\piz$) decay, the origin of the photons from the $\piz$ is taken from 
the common vertex found by fitting the three tracks in the signal hemisphere.

The respective mass windows for the reconstructed $\eta$ candidates are:
\begin{itemize}
\item  $0.515\gevcc < M(\eta \to \gamma\gamma)       < 0.565\gevcc$,
\item  $0.115\gevcc < M(\piz \to \gamma\gamma)       < 0.150\gevcc$,
\item  $0.537\gevcc < M(\eta \to \pi^{+}\pi^{-}\piz) < 0.558\gevcc$.
\end{itemize}

Finally, we kinematically fit for the momentum of the $\eta$ and the $\piz$ daughters after applying $\eta$ and $\piz$ mass constraints
on their respective daughters. We then combine the signal muon track and the $\eta$ candidates to form $\tau^{-}$ candidates. 
The signal $\tau^- \to \mu^- \eta$ decays are identified by two kinematic variables: the beam-energy constrained $\tau$ mass (\mec) and 
$\Delta E = E_{\mu} + E_{\eta} - \roots/2$, where $E_\mu$ and $E_{\eta}$ are the energy of the $\mu$ and 
the $\eta$ candidate in the \CM frame. These two variables are independent apart from small correlations arising 
from initial and final state radiation. 

The mean and standard deviation of the \mec and \DeltaE distributions near their statistical maxima for the reconstructed 
$\tau^{-} \to \mu^{-}\eta ~(\eta \to \gamma\gamma)$ MC signal events are:
$\langle \mec \rangle$ = 1777.9\mevcc, $\sigma(\mec)$ = 8.25\mevcc,
$\langle \DeltaE \rangle$ = $-$13.4\mev, $\sigma(\DeltaE)$ = 40.8\mev,
whereas for $\tau^{-} \to \mu^{-}\eta ~(\eta \to \pi^{+}\pi^{-}\piz)$ MC signal events the corresponding quantities are: 
$\langle \mec \rangle$ = 1777.7\mevcc, $\sigma(\mec)$ = 5.6\mevcc, 
$\langle \DeltaE \rangle$ = $-$7.1\mev, $\sigma(\DeltaE)$ = 31.0\mev.
The shift from zero in $\langle \DeltaE \rangle$ comes from the leakage of the measured photon energy from deposits in the EMC.
We do not look at the data events within a $\pm$ 3$\sigma$ rectangular region centered at
($\langle \mec \rangle$, $\langle \DeltaE \rangle$) in the \mec vs. \DeltaE plane,
until completing all optimization and systematic studies of the selection criteria.

The \mtau candidate is considered for further analysis if the signal lepton track is consistent with being identified as a muon, 
and not an electron or a kaon, using a set of particle identification (PID) criteria
using DCH, EMC, DIRC and IFR information, which has a muon identification efficiency of about 85\% and 65\% 
inside the polar angle coverages of $[17^\circ, 57^\circ]$ and $[57^\circ, 155^\circ]$, respectively, 
for tracks with momentum greater than 0.5\gevc.
This provides the correct association of the reconstructed particles having the \mtau as their mother 
for $99.8\%$ of selected signal MC events inside a $\pm$ 5$\sigma$ rectangular region in \mec vs. \DeltaE plane 
for both the decay channels.

To reduce backgrounds from $\epem \to \mumu$ processes, we require the total \CM momentum of 
all the tracks observed in the DCH and unassociated energy deposits in the EMC on the tag-side hemisphere
(defined to be the one opposite to the signal-side hemisphere) to be less than 4.75\gevc for both the channels. 

Other non-$\tau$ backgrounds are suppressed by requiring the polar angle 
of the missing momentum ($\theta_{miss}$) associated with the neutrinos in the event
to lie within the detector acceptance $(-0.76 < \cos \theta_{miss} < 0.92)$,
and the scaled missing \CM transverse momentum relative to the beam axis $(p^{T}_{miss}/\sqrt{s})$ to be greater than 0.082.

A tag-side hemisphere containing a single track is classified as e$-$tag, $\mu-$tag or  h$-$tag
if the track is exclusively identified as an electron (e$-$tag), as a muon ($\mu-$tag), or as neither (h$-$tag)
and if the total neutral \CM energy in the hemisphere is no more than 200\mev.
If the total neutral \CM energy in the hemisphere is more than 200\mev and the track is neither an electron nor a muon
it is classified as $\rho-$tag. If the tag-side contains three tracks, the event is classified as a 3h-tag for the
$\tau^{-} \to \mu^{-}\eta~(\eta \to \gamma\gamma)$ search, whereas for the 3-prong signal decay channel of $\eta$ we
use only 1-prong decays of the \mtau in the tag-side hemisphere.

We assume that the $\tau^{-} \to \mu^{-}\eta$ decay fully reconstructs 
the direction of the $\tau^{+}$ in the hemisphere opposite the signal candidate.
We then calculate the invariant mass squared (\Mnu) of the unobserved particle, assumed to be $\nu(\mathrm{s})$,
from the $\tau^{+}$ decay, assuming that the energy of the $\tau^{+}$ is given by the beam energy. 

A cut of $-1.0\gevccgevcc < \Mnu < 2.5\gevccgevcc$ is applied to the e$-$tag, $\mu-$tag and h$-$tag
in the search for the $\tau^{-} \to \mu^{-}\eta~(\eta \to \gamma\gamma)$ decay.
For the $\rho-$tag, we require $-1.0\gevccgevcc < \Mnu < 1.0\gevccgevcc$ and for the 3h$-$tag, 
we require $-0.2\gevccgevcc < \Mnu < 1.0\gevccgevcc$ in the $\eta \to \gamma\gamma$ decay mode. 
For $\eta \to \pi^+\pi^-\piz$ mode, we require $-1.25\gevccgevcc < \Mnu < 2.5\gevccgevcc$ for all tags.
In addition to the above requirements, we apply cuts on the invariant mass on the tag-side (\tagmass),
calculated from the tracks observed in the DCH and unassociated neutral energy deposits in the EMC,
to be less than $0.4\gevcc$ for the h$-$tag and $0.6\gevcc < \tagmass < 1.35\gevcc$ for the $\rho-$tag.

The cuts on the different variables have been applied so as to minimize the expected upper limit 
at 90\% c.l.~\cite{Feldman:1997qc} on the branching ratio of the signal $\tau^- \to \mu^-\eta$ decay,
in a background only hypothesis, estimated using Monte Carlo (MC) simulation of background events.

\begin{figure}[!h]
  \begin{center}
    {\epsfig{file=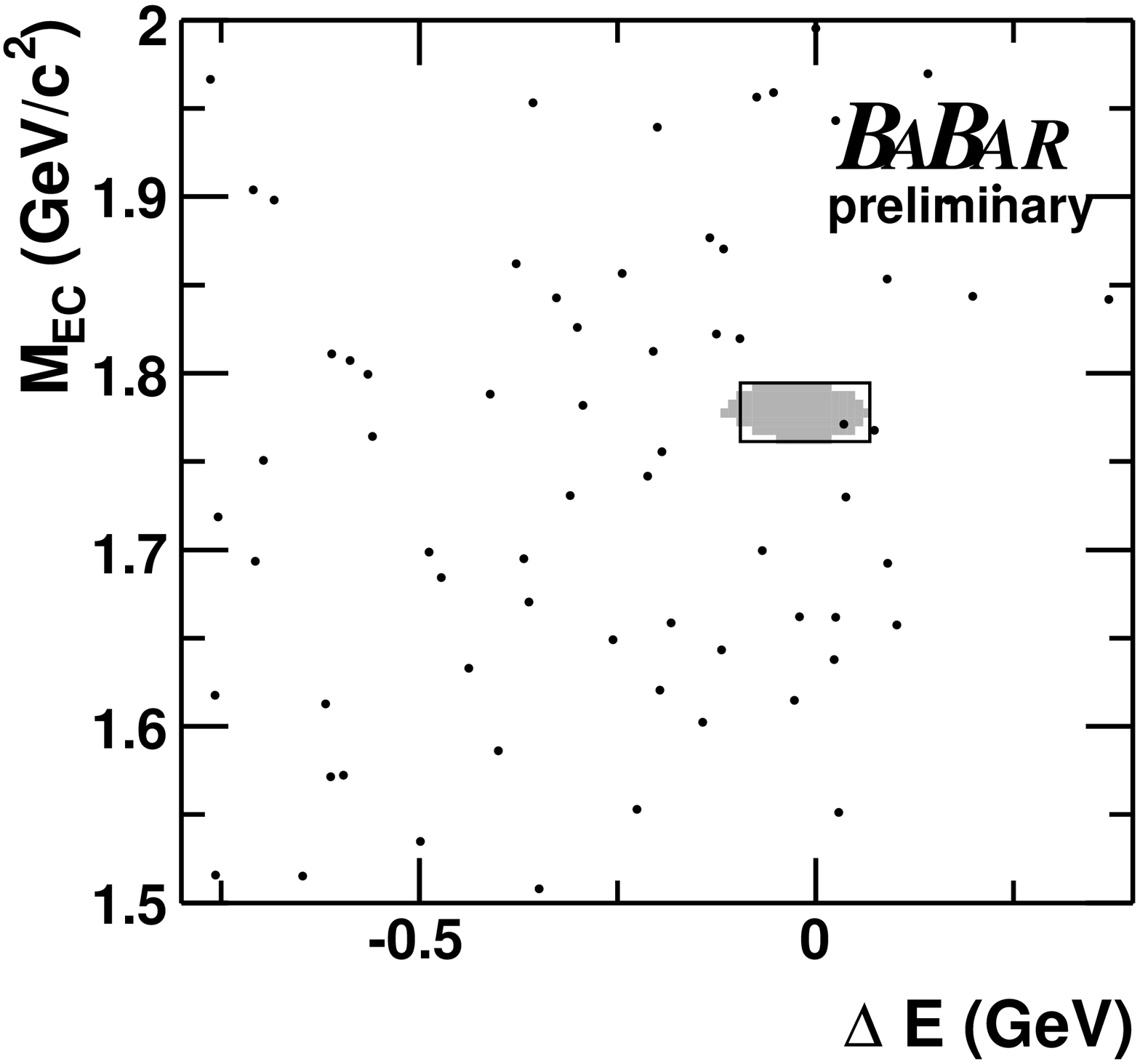,width=.49\textwidth}}
    {\epsfig{file=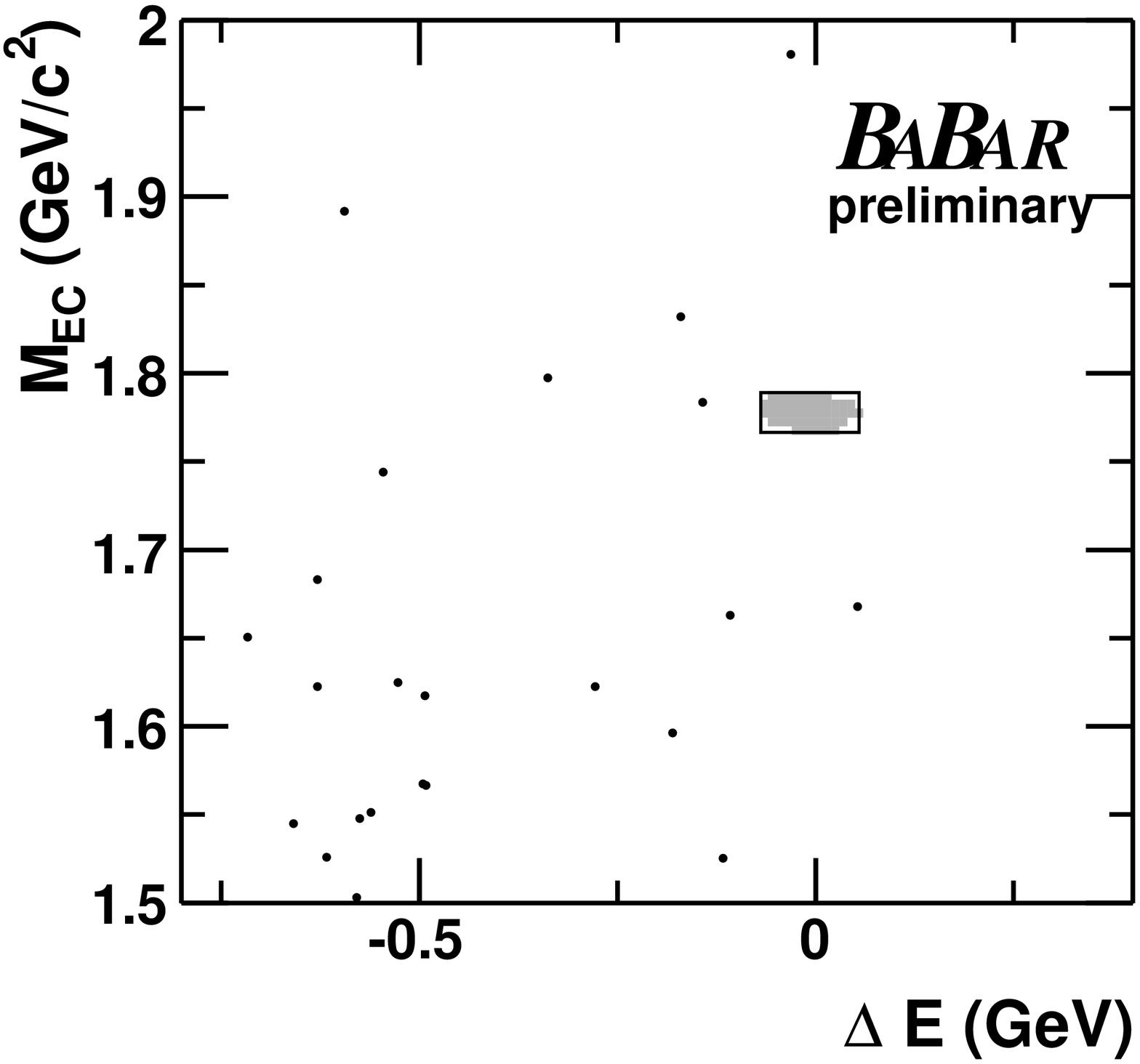,width=.49\textwidth}}
    \caption{The \DeltaE and \mec  ditributions for the data (shown as dots) after all the selection criteria 
             have been applied, in the search for the $\tau^{-} \to \mu^{-}\eta ~(\eta \to \gamma\gamma)$ decay (left)
             and the $\tau^{-} \to \mu^{-}\eta ~(\eta \to \pi^{+}\pi^{-}\piz)$ decay (right).
             The shaded region contains 70\% of the selected signal MC events inside the Grand Signal Box for both modes.
             The $\pm$2$\sigma$ signal box is also overlayed.}
    \label{mec_deltae}
  \end{center}
\end{figure}

After this selection, 10.08\% and 5.43\% of the $\tau^{-} \to \mu^{-}\eta~(\eta \to \gamma\gamma)$ and
$\tau^{-} \to \mu^{-}\eta~(\eta \to \pi^+\pi^-\piz)$ MC signal events survive within
a Grand Signal Box (GSB) region defined as: \mec $\in$ $[1.5, 2.0]$ \gevcc, \DeltaE $\in$ $[-0.8, 0.4]$ \gev.
The data distribution of \mec and \DeltaE inside the GSB is plotted as dots in Figure~\ref{mec_deltae}.
About 70\% of the selected signal MC events inside the GSB lie
within a $\pm$2$\sigma$ rectangular region in the \mec vs. \DeltaE plane.
These MC events are shown by the shaded region in Figure~\ref{mec_deltae}.

\subsection{Background estimation}

The GSB regions excluding the $\pm$3$\sigma$ blinded region contain 65 and 23 data events,
while the luminosity-normalized sum of the MC backgrounds yield ($74.6\pm9.0$) and ($34.3\pm5.2$) events for the 2 channels respectively.
For the $\tau^{-} \to \mu^{-}\eta~(\eta \to \gamma\gamma)$ channel,
24\%, 62\% and 9\% of the MC background events have a true $\mu$ only,
a true $\eta$ candidate only and both true $\mu$ and true $\eta$ candidates, respectively.
For the $\tau^{-} \to \mu^{-}\eta~(\eta \to \pi^+\pi^-\piz)$ channel,
4\%, 37\% and 4\% of the MC background events have a true $\mu$ only,
a true $\eta$ candidate only and both true $\mu$ and true $\eta$ candidates, respectively.
The major source of backgrounds come from either a mis-identification of a pion track as a muon candidate
or the in-efficiency in $\eta$ and $\piz$ reconstruction.

\begin{figure}[!h]
  \begin{center}
    {\epsfig{file=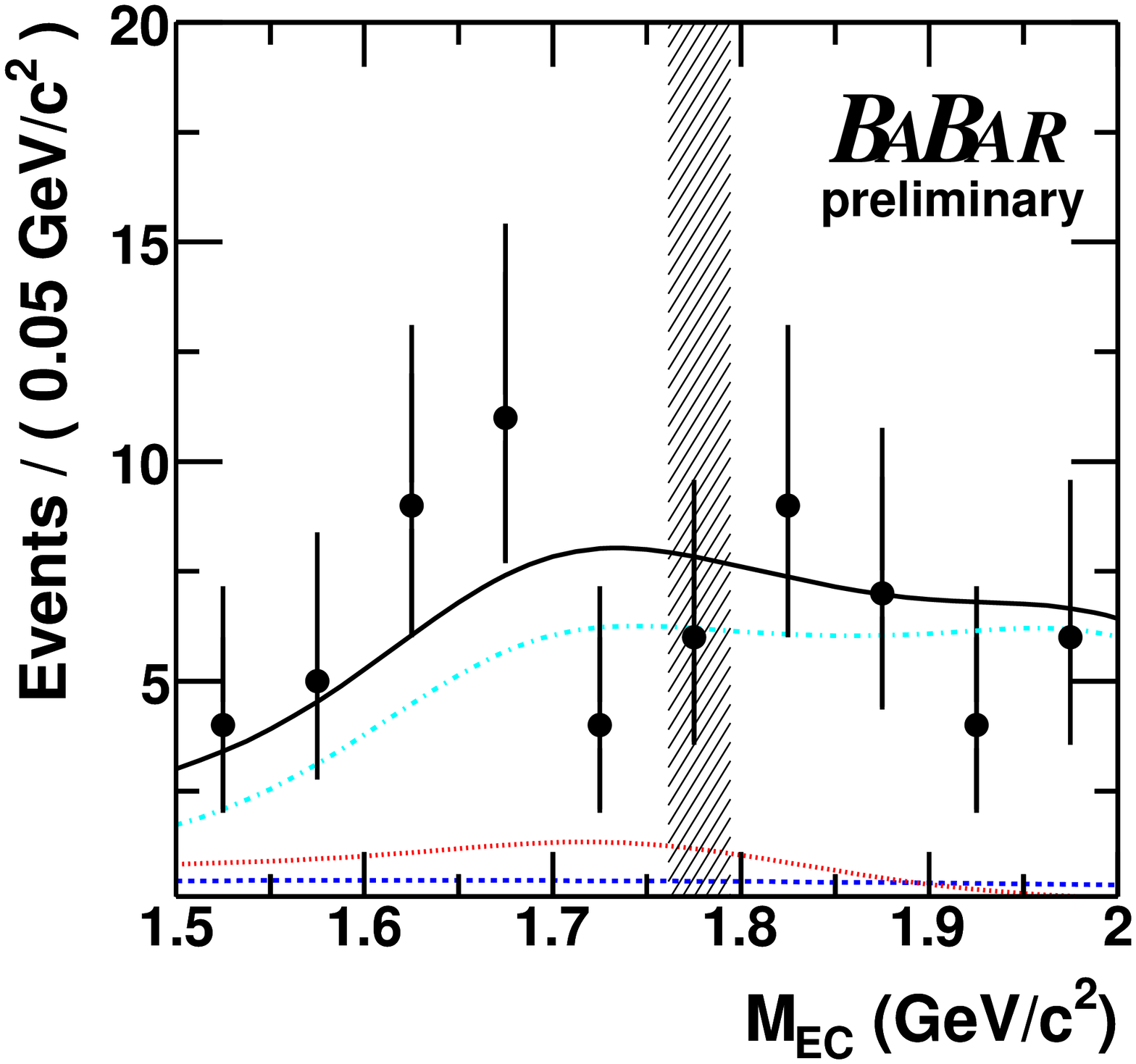,width=.49\textwidth}}
    {\epsfig{file=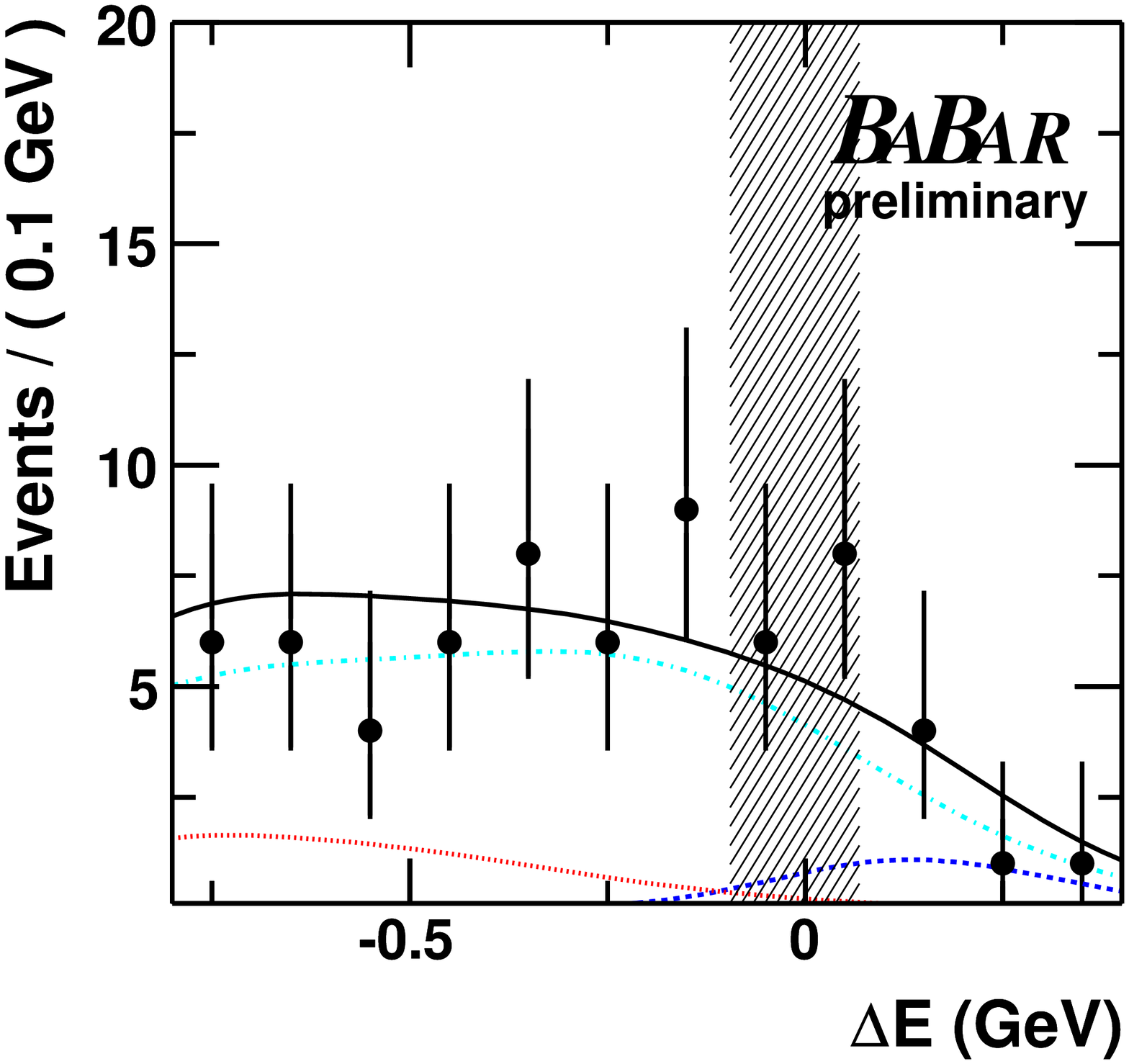,width=.49\textwidth}}
    {\epsfig{file=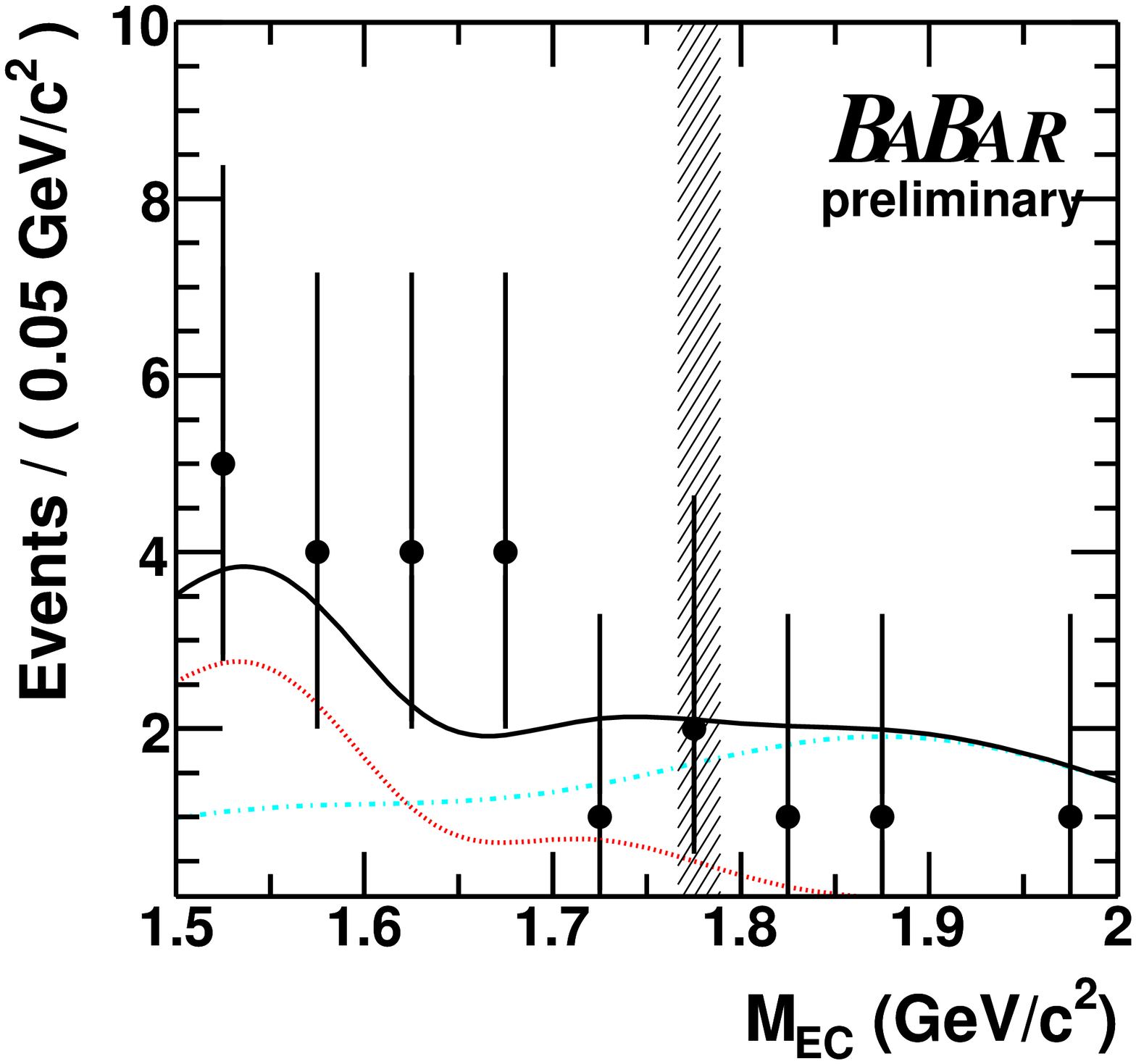,width=.49\textwidth}}
    {\epsfig{file=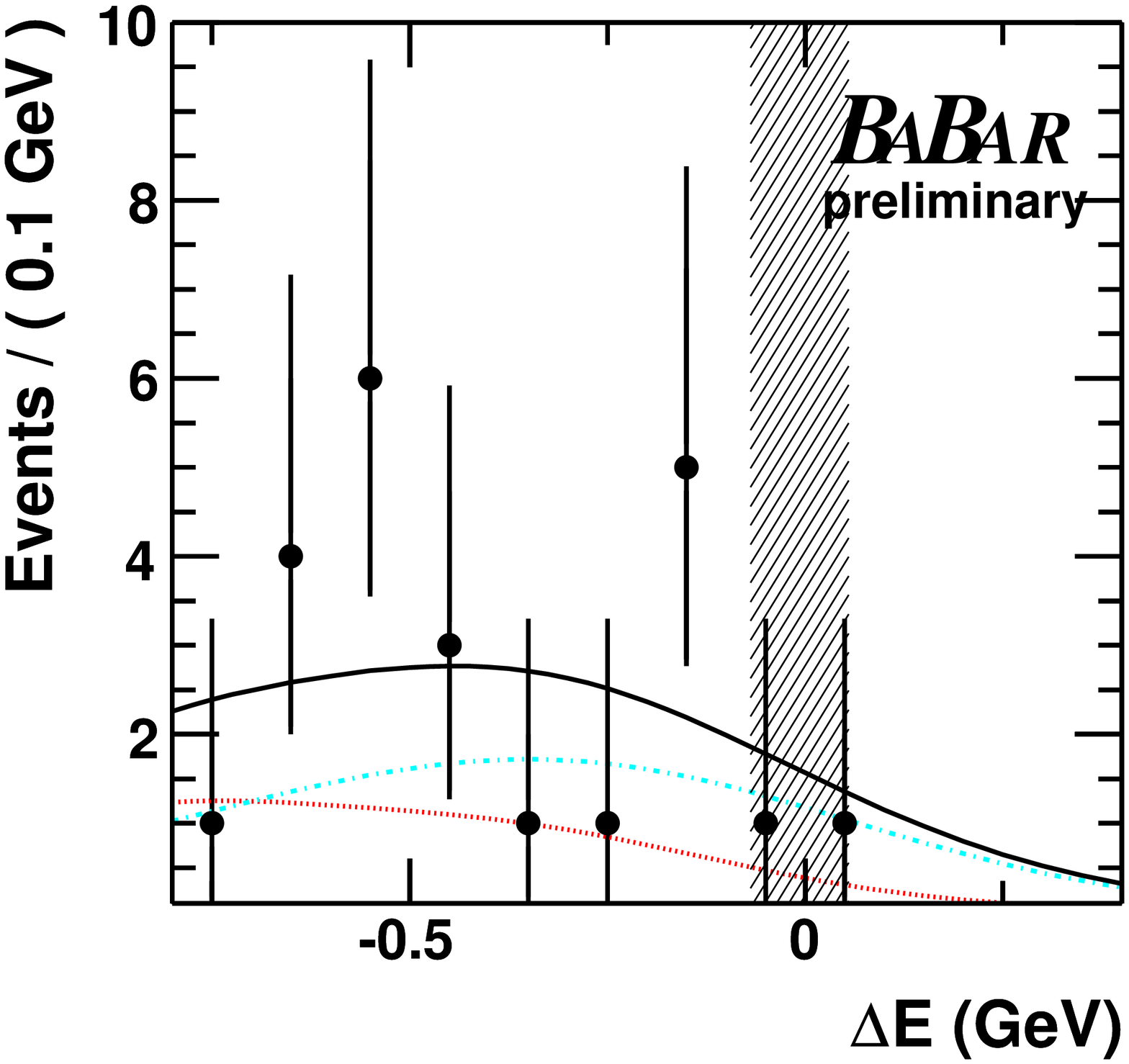,width=.49\textwidth}}
    \caption{The fit to the data distributions of \mec  and \DeltaE variables
	     in the GSB regions excluding the $\pm$ 3$\sigma$ blinded region
             in the search for the $\tau^{-} \to \mu^{-}\eta ~(\eta \to \gamma\gamma)$ decay (top row)
             and for the $\tau^{-} \to \mu^{-}\eta ~(\eta \to \pi^{+}\pi^{-}\piz)$ decay (bottom row)
             are shown along with the data distributions (as dots) and the background MC components 
             ($\mumu$ in dashed, $\tautau$ in dotted and uds in dashed-dotted).
             The shaded region is the $\pm$2$\sigma$ signal box for both modes.
             }
    \label{final_fit}
  \end{center}
\end{figure}

The number of expected backgrounds in the signal box is extracted from 
an un-binned maximum likelihood fit to the distributions of the \mec and \DeltaE variables for the data events 
inside the non-blinded parts of the GSB region, while the shape of the distributions have been obtained from MC.
The number of background events ($N^{data}_{2\sigma}$) inside the $\pm$2$\sigma$ signal box is estimated as:
$$ N^{data}_{2\sigma} = \frac{\int_{2\sigma} PDF_{tot}}
                        {\int_{GSB - 3\sigma} PDF_{tot}} 
                        \times N^{data}_{GSB-3\sigma}$$
where $\int_{2\sigma}PDF_{tot}$ and $\int_{GSB- 3\sigma}PDF_{tot}$ are the probability density functions (PDFs) 
integrated over the the signal box and the non-blinded parts of the GSB regions,
$N^{data}_{GSB-3\sigma}$ are the number of data events in the non-blinded parts of the GSB region, and the $PDF_{tot}$ is defined as:
$$PDF_{tot} = (f_{\mu}\times PDF_{\mu}) + (f_{\tau}\times PDF_{\tau}) + ([1-f_{\mu}-f_{\tau}]\times PDF_{uds})$$
where the $f_{\mu}$ and $f_{\tau}$ are the fractions of $\mu^+\mu^-$ and $\tau^+\tau^-$ background contributions. 
The $PDF_{\mu}$, $PDF_{\tau}$ and $PDF_{uds}$ are non-parametric PDFs~\cite{keys} 
obtained from the respective background MC events 
after applying the same final selection criteria.
The corresponding projections of the PDFs obtained from MC for the individual components and the total fit to the data
in the GSB region excluding the $\pm$3$\sigma$ blinded region for \mec and \DeltaE variables are shown in 
Figure~\ref{final_fit}, along with the data points. The $\pm$2$\sigma$ signal box is indicated by hatches. 

The number of background events from the fit to the data in the non-blinded parts of the GSB region
for $\tau^{-} \to \mu^{-}\eta ~(\eta \to \gamma\gamma)$ and 
$\tau^{-} \to \mu^{-}\eta ~(\eta \to \pi^{+}\pi^{-}\piz)$ searches are (0.64$\pm$0.08) and  (0.07$\pm$0.02) events, respectively.
The observed and the expected number of background events inside the neighbouring boxes in the non-blinded parts of the GSB 
are shown in Table~\ref{tab:sidebox_mode2} and Table~\ref{tab:sidebox_mode4} for the respective channels.

As a cross-check, we also estimate the background assuming a uniform distribution over the GSB in \mec vs. \DeltaE plane.
This simple interpolation gives the number of backgrounds in signal box to be (0.60$\pm$0.07) and (0.10$\pm$0.02) for
$\tau^{-} \to \mu^{-}\eta ~(\eta \to \gamma\gamma)$ and $\tau^{-} \to \mu^{-}\eta ~(\eta \to \pi^{+}\pi^{-}\piz)$, respectively.

\begin{table}[!h]
\begin{center}
\renewcommand{\arraystretch}{1.10}
\begin{tabular}{c|l|l|l|l|l} 
\hline\hline
$\tau^{-} \to \mu^{-}\eta ~(\eta \to \gamma\gamma)$ 
                        & $(5-3)\sigma$ & $(7-5)\sigma$ & $(9-7)\sigma$ & $(11-9)\sigma$ & $(11-3)\sigma$ \\\hline 
\# of observed events   & 4             &  3            & 5             & 7              & 19             \\
\# of expected events   & 2.5$\pm$0.3   &  3.7$\pm$0.5  & 4.8$\pm$0.6   & 5.7$\pm$0.7    & 16.7$\pm$2.1   \\ 
\hline\hline
\end{tabular}
\end{center}
\caption[]
{The number of observed and expected data events inside the $5\sigma-3\sigma$, $7\sigma-5\sigma$, $9\sigma-7\sigma$, 
$11\sigma-9\sigma$ and $11\sigma-3\sigma$ neighboring boxes in the \mec vs. \DeltaE for the search for the
decay $\tau^{-} \to \mu^{-}\eta ~(\eta \to \gamma\gamma)$.}
\label{tab:sidebox_mode2}
\end{table}

\begin{table}[!h]
\begin{center}
\renewcommand{\arraystretch}{1.10}
\begin{tabular}{c|l|l|l|l|l} 
\hline\hline
$\tau^{-} \to \mu^{-}\eta ~(\eta \to \pi^+\pi^-\piz)$ 
                         & $(5-3)\sigma$ & $(7-5)\sigma$ & $(9-7)\sigma$ & $(11-9)\sigma$ & $(11-3)\sigma$\\\hline 
\# of observed events    & 1             &  0            & 0             & 2              & 3             \\
\# of expected events    & 0.3$\pm$0.1   &  0.4$\pm$0.1  & 0.5$\pm$0.1   & 0.7$\pm$0.1    & 1.9$\pm$0.4   \\ 
\hline\hline
\end{tabular}
\end{center}
\caption[]
{The number of observed and expected data events inside the $5\sigma-3\sigma$, $7\sigma-5\sigma$, $9\sigma-7\sigma$, 
$11\sigma-9\sigma$ and   $11\sigma-3\sigma$ neighboring boxes in the \mec vs. \DeltaE plane for the search for the 
decay $\tau^{-} \to \mu^{-}\eta ~(\eta \to \pi^{+}\pi^{-}\piz)$.}
\label{tab:sidebox_mode4}
\end{table}

\section{SYSTEMATIC STUDIES}
\label{sec:Systematics}

Table \ref{tblSystematics} is a summary of the relative systematic uncertainties in the signal efficiency estimation.
The largest systematic uncertainities are due to the signal track momentum and the photon energy scale and resolution, 
which are the tracking and calorimetry errors introduced by applying the final cut on the $\pm$2$\sigma$ signal region.
In order to assess the systematic errors associated with these scale and resolution uncertainties, 
the peak and resolution of \mec as well as the \DeltaE are varied. 
The errors associated with the modeling of each election variable is estimated from the 
relative change in signal efficiency when varying the cut by the Data-MC difference in the mean of that variable.

Other sources of systematic uncertainities include those arising from trigger and filter efficiencies,
tracking and neutral energy reconstruction efficiencies, the PID error associated with signal muon track identification,
beam energy scale and spread, luminosity estimation and cross-section of the $\epem\to\tautau$ process.
 
As we use 2.3 million MC signal events, the contribution to the
uncertainty arising from signal MC statistics is negligible.

All contributions to the systematic uncertainties are added in quadrature to give a total relative systematic
uncertainty of $7.5\%$ and $8.6\%$ for the $\tau^{-} \to \mu^{-}\eta ~(\eta \to \gamma\gamma)$ and 
the $\tau^{-} \to \mu^{-}\eta ~(\eta \to \pi^{+}\pi^{-}\piz)$ channels respectively.

\begin{table}[htb]
\begin{center}
\renewcommand{\arraystretch}{1.10}
\begin{tabular}{lll}
\hline \hline
Uncertainty ($\delta\eff/\eff$)                       & $\%$        &  $\%$    \\
                                                      &($\eta \to \gamma\gamma$) & ($\eta \to \pi^{+}\pi^{-}\piz$)\\
\hline			                 		   
Trigger and filter efficiency                         & $1.0$       & $0.4$     \\
Tracking efficiency                                   & $1.3$       & $2.2$     \\
Neutral energy reconstruction efficiency              & $3.3$       & $3.3$     \\
PID error associated with signal muon track           & $3.0$       & $2.7$     \\
Modeling of the selection variables                   & $3.9$       & $3.4$     \\
Signal track and photon energy scale and resolution   & $3.7$       & $5.8$     \\
Beam energy scale and spread                          & $0.4$       & $0.5$     \\
Luminosity and \tautau production cross-section       & $2.3$       & $2.3$     \\
\hline			                        		       
Total                                                 & $7.5$       & $8.6$     \\

\hline \hline
\end{tabular}
\end{center}
\caption{Summary of the relative systematic uncertainties (in \%) of the signal efficiency for both decay modes 
         $\tau^{-} \to \mu^{-}\eta ~(\eta \to \gamma\gamma)$ and 
         $\tau^{-} \to \mu^{-}\eta ~(\eta \to \pi^{+}\pi^{-}\piz)$.}
\label{tblSystematics}
\end{table}

The dominant contribution to the uncertainity of background estimation arises from the statistical error on the number of data events 
surviving the final selection inside the non-blinded parts of the GSB, and from the variation of the fitted 
$f_{\mu}$ and $f_{\tau}$ within $\pm$1$\sigma$ from the fit to the data.
A small contribution of 1.9\% and 0.9\% on the relative uncertainities from a bias in the fit 
arising from blinding of the $\pm$3$\sigma$ rectangular region in the \mec vs. \DeltaE plane
during estimation of PDFs from the MC background samples has been included in the above-mentioned estimate.

\section{RESULTS}
\label{sec:Physics}

The upper limit of $\tau^-\to\mu^-\eta$ is calculated using $\BR^{90}_{UL}=N^{90}_{UL}/(2\L\sigma_{\tau\tau}\BR\eff)$,
where $N^{90}_{UL}$ is the 90\% c.l. upper limit on the number of signal events expected within the $\pm2\sigma$ box
in \mec vs. \DeltaE plane, \eff is the reconstruction efficiency and \BR\ is the  branching ratio of the $\eta$ decay modes
under consideration.

To obtain a combined upper limit, we add the signal efficiencies for the individual channels
weighted by their respective branching ratios using the formula:
$$\BR\eff = (\BR_1 \times \eff_1 + \BR_2 \times \eff_2)$$
where $\eff_{1}$ = (7.03$\pm$0.53)\%, $\eff_{2}$ = (3.67$\pm$0.32)\%
are the signal reconstruction efficiencies for the two decay modes
$\tau^- \to \mu^-\eta ~(\eta \to \gamma\gamma)$ and $\tau^- \to \mu^-\eta ~(\eta \to \pi^+\pi^-\piz)$,
and $\BR_1$, $\BR_2$ are the branching ratios of $\eta \to \gamma\gamma$ and 
$\eta \to \pi^+\pi^-\piz (\piz \to \gamma\gamma)$, respectively.
Thus, for the decay modes combined, taking into account the errors from similar sources as correlated,
we have a combined efficiency for reconstructing $\tau^{-} \to \mu^{-}\eta$ of $\BR\eff$ = (3.59 $\pm$ 0.41)\%.
This corresponds to a 11.4\% relative systematic uncertainty on the total selection efficiency.

We sum the backgrounds inside the $\pm$2$\sigma$ signal box from the above two channels to obtain a background expectation 
of (0.71 $\pm$ 0.08) events for the combined upper limit calculation.
Inside the $\pm$2$\sigma$ signal box, we observe only one event in the search for the decay 
$\tau^- \to \mu^-\eta ~(\eta \to \gamma\gamma)$,
and none in the search of the decay $\tau^- \to \mu^-\eta ~(\eta \to \pi^+\pi^-\piz)$.

The  limit  is calculated including all uncertainties
using the technique of Cousins and Highland~\cite{Cousins:1992qz}
following the implementation of Barlow~\cite{Barlow:2002bk}.
In this technique,  MC samples are generated according to a Poisson distribution with mean
$(s + b)$ where the background, $b$,  and signal, $s$, are each drawn randomly from Gaussian distributions
describing their respective PDFs. The values of the mean and standard deviation
for the background Gaussian are 0.71 and 0.08 events, respectively. The mean of the signal Gaussian
is $(2\L\sigma_{\tau\tau}\BR_{UL}\eff)$  and the standard deviation is the error on $(2\L\sigma_{\tau\tau}\BR_{UL}\eff)$.
The branching ratio, $\BR_{UL}$, is varied until we find a value for which 10\% of the sample yields a number of  events
less than the one event observed in the data.
At 90\% c.l. this procedure gives an observed upper limit for $\BR(\tau^{-} \to \mu^{-}\eta)$ to be 1.6\tenseven
including effects of systematic uncertainties on the signal efficiency and background estimation 
for both the decay modes combined. Averaging the number of observed events along with its poisson fluctuation,
the expected upper limit for $\BR(\tau^{-} \to \mu^{-}\eta)$ at 90\% c.l. is 1.4\tenseven.

The $\eta$ branching ratio, the selection efficiency, the number of expected and observed background events
and the observed upper limit at 90\% c.l. with the systematic uncertainities are shown in Table~\ref{tab:final} 
for the two modes separately and combined.

\begin{table}[!h]
\begin{center}
\renewcommand{\arraystretch}{1.10}
\begin{tabular}{l          |c               |l                  |  l             |      c   | l       } \hline\hline
Decay modes                & Branching Ratio & Efficiency & \multicolumn{2}{|c|}{Background events}  & Upper Limit\\ \cline{4-5}
                           & \multicolumn{1}{|c|}{of $\eta$(\%)} & \multicolumn{1}{|c|}{(\%)} & Expected  &  Observed & (@90\% c.l.)\\\hline
$\tau^-\to\mu^-\eta~(\eta\to\gaga)$      & 39.42$\pm$0.26 & 7.03$\pm$0.53      & 0.64$\pm$0.08  &   1      & 2.1\tenseven \\ \hline
$\tau^-\to\mu^-\eta~(\eta\to\pi^+\pi^-\piz)$ & 22.32$\pm$0.40 & 3.67$\pm$0.32      & 0.07$\pm$0.02  &   0      & 4.9\tenseven \\ \hline
$\tau^-\to\mu^-\eta$                     & \multicolumn{2}{|c|}{3.59$\pm$0.41} & 0.71$\pm$0.08  &   1      & 1.6\tenseven \\ \hline\hline
\end{tabular}
\end{center}
\caption[]
{The $\eta$ branching ratio, the selection efficiency, the number of expected and observed background events and the observed upper limit
for the two modes separately and combined with systematic uncertainity.}
\label{tab:final}
\end{table}

\section{SUMMARY}
\label{sec:Summary}

The search for the SM forbidden decay $\tau^- \to \mu^-\eta$ is performed using 314.5 \invfb data. 
We find one event consistent with the signal signature for an expected background of (0.71$\pm$0.08) events.
A 90\% c.l. upper limit on the branching ratio is calculated to be 1.6\tenseven
including effects of systematic uncertainties on the signal efficiency and background estimation.

\section{ACKNOWLEDGMENTS}
\label{sec:Acknowledgments}

% Standard acknowledgments paragraph; must always be included.
We are grateful for the 
extraordinary contributions of our \pep2\ colleagues in
achieving the excellent luminosity and machine conditions
that have made this work possible.
The success of this project also relies critically on the 
expertise and dedication of the computing organizations that 
support \babar.
The collaborating institutions wish to thank 
SLAC for its support and the kind hospitality extended to them. 
This work is supported by the
US Department of Energy
and National Science Foundation, the
Natural Sciences and Engineering Research Council (Canada),
Institute of High Energy Physics (China), the
Commissariat \`a l'Energie Atomique and
Institut National de Physique Nucl\'eaire et de Physique des Particules
(France), the
Bundesministerium f\"ur Bildung und Forschung and
Deutsche Forschungsgemeinschaft
(Germany), the
Istituto Nazionale di Fisica Nucleare (Italy),
the Foundation for Fundamental Research on Matter (The Netherlands),
the Research Council of Norway, the
Ministry of Science and Technology of the Russian Federation, and the
Particle Physics and Astronomy Research Council (United Kingdom). 
Individuals have received support from 
the Marie-Curie IEF program (European Union) and
the A. P. Sloan Foundation.


\begin{thebibliography}{99}
\bibitem{conjugate}
Charge conjugate mode for all the decays are implied throughout this paper unless otherwise stated.
\bibitem{LAB}
All measured quantities are quoted in the laboratory frame,
unless otherwise mentioned.
\end{thebibliography}
\end{document}